\documentclass[12pt,a4paper]{article}
 
\usepackage{amsmath,amssymb,amsthm}
\usepackage{graphicx}
\usepackage{hyperref}
\usepackage{cite}
\usepackage{physics}
\usepackage[margin=2.5cm]{geometry}
\usepackage{xcolor}
\usepackage{float}
\usepackage{braket}

\newtheorem{remark}{Remark}
 
\newcommand{\DMO}{\mathrm{DMO}}
\newcommand{\IDMO}{\mathrm{IDMO}}

\begin{document}

\title{The Inverted Dirac-Moshinsky Oscillator in $(1+1)$ Dimensions}
 
\author{Kevin Hernández$^{1}$ \and Marcos Iraheta-Orellana$^{1}$ \and Wiliam Larín-Escobar$^{2}$}
 
\date{%
  $^{1}$Escuela de Física, Facultad de Ciencias Naturales y Matemáticas, Universidad de El Salvador, Final de Av. Mártires y Héroes del 30 de julio, San Salvador, El Salvador\\
  $^{2}$Far Eastern Federal University, Vladivostok, Russia\\
  [6pt]
  \small\texttt{kevinhernandezbel@hotmail.com}\\[6pt]
  \today
}
 
\maketitle
% ============================================================
\begin{abstract}
We derive and analyze the exact solutions of the inverted
Dirac-Moshinsky oscillator (IDMO) in $(1+1)$ dimensions,
obtained from the standard model via the substitution
$p \to p + im\omega\beta x$.
The upper spinor component satisfies a Weber equation with
complex spectral parameter $\lambda = (E^2-m^2)/(2m\omega)+i/2$,
whose solutions are parabolic cylinder functions $D_\nu(\xi)$
with complex order $\nu = \lambda - 1/2$.
The physical spectrum is purely continuous ($|E|>m$), with
no discrete bound states.
Three normalization schemes are developed, and the discrete
Gamow resonances at $E_n^\pm = \pm\sqrt{m^2+(2n+1)m\omega-im\omega}$
are identified as poles of the resolvent.
The negative-energy sector describes antiparticle
anti-resonances whose positive imaginary part signals
vacuum instability and spontaneous pair production,
 analogous to the Schwinger effect.
The algebraic structure is governed by the principal series
of $SU(1,1)$, and the Hamiltonian is $\mathcal{PT}$-symmetric
with unbroken symmetry for $|E|>m$.
\end{abstract}
 
\noindent\textbf{Keywords:}
Dirac-Moshinsky oscillator, inverted oscillator,
parabolic cylinder functions, continuous spectrum,
Rigged Hilbert Space, $\mathcal{PT}$-symmetry.

% ============================================================
\section{Introduction}
\label{sec:intro}
 
The Dirac-Moshinsky oscillator (DMO), introduced by Moshinsky
and Szczepaniak in 1989~\cite{Moshinsky1989}, is among the
most celebrated exactly solvable models in relativistic quantum
mechanics.
Defined through the nonminimal substitution
$\mathbf{p} \to \mathbf{p} - im\omega\beta\mathbf{r}$
in the free Dirac equation, the DMO produces, in the
nonrelativistic limit, the conventional harmonic oscillator
supplemented by a strong spin-orbit coupling term.
Its exact solvability, rich algebraic structure, and
connections to diverse areas of physics have motivated
an extensive literature spanning supersymmetric quantum
mechanics~\cite{Sadurni2011}, the Jaynes-Cummings model
of quantum optics~\cite{Bermudez2007},
graphene physics~\cite{Sadurni2011}, quantum chromodynamics,
and topological defects.
 
The mathematical structure of the standard DMO is thoroughly
understood: its spectrum is discrete, the Hamiltonian is
essentially self-adjoint, and its eigenfunctions form a
complete orthonormal basis in $L^2(\mathbb{R})$.
The underlying symmetry algebra is $SU(1,1) \oplus SO(2,1)$,
which accounts for the characteristic supersymmetric
pattern of the energy spectrum~\cite{Sadurni2011}.
 
In contrast, the \emph{inverted} Dirac-Moshinsky oscillator
(IDMO), obtained through the replacement $\omega \to i\omega$
(or equivalently $\mathbf{p} \to \mathbf{p} + im\omega\beta\mathbf{r}$),
has received considerably less attention despite its physical
relevance.
At the level of the effective potential, this analytic
continuation converts the confining harmonic interaction into
an inverted harmonic oscillator, radically altering the
spectral and dynamical properties of the system.
Instead of discrete bound states, one encounters resonant
solutions, non-normalizable states, and a continuous spectrum,
phenomena that require tools beyond the standard Hilbert
space formulation.
 
The inverted harmonic oscillator occupies a central role in
modern theoretical physics.
It appears in the description of unstable quantum
systems~\cite{Barton1986}, tunneling phenomena, cosmological
particle production~\cite{Parker1969}, black-hole
physics~\cite{Hawking1975}, quantum chaos and
complexity~\cite{Hashimoto2020}, and the theory of Gamow
vectors.
In all these contexts, the non-normalizable eigenstates of the
inverted potential require the framework of Rigged Hilbert
Spaces (RHS), or Gel'fand triplets~\cite{deLaMadrid2002},
\begin{equation}
  \Phi \subset \mathcal{H} \subset \Phi^\times \,,
  \label{eq:RHS}
\end{equation}
where generalized eigenvectors may be consistently defined
as continuous functionals over an appropriate test space
$\Phi$.
 
The purpose of the present work is to provide a complete,
self-contained derivation of the IDMO in $(1+1)$ dimensions,
establishing its exact solutions, spectral properties, and
algebraic structure.
We derive the parabolic cylinder equation satisfied by the
upper spinor component, obtain the general solution in terms
of $D_\nu(z)$ with complex order $\nu$, and discuss the
spectral interpretation within the RHS framework.
The results presented here lay the groundwork for further
investigations of the IDMO in the context of
$\mathcal{PT}$-symmetric quantum mechanics~\cite{Bender1998,
Bender2007}, non-Hermitian field theories, and relativistic
scattering in external fields.
 
The paper is organized as follows.
Section~\ref{sec:hamiltonian} defines the IDMO Hamiltonian
in $(1+1)$ dimensions.
Section~\ref{sec:eigenvalue} sets up the eigenvalue problem.
Section~\ref{sec:second_order} derives the second-order
differential equation for the upper component.
Section~\ref{sec:change_variable} reduces it to the Weber
equation.
Section~\ref{sec:solution} presents the exact solution.
Section~\ref{sec:spectrum} analyzes the spectral properties.
Section 8 derives the energy values, normalization schemes,
and explicit wave functions.
Section 9 discusses the algebraic structure.
Section 10 summarizes the results.
 
% ============================================================
\section{The Hamiltonian}
\label{sec:hamiltonian}
 
In $(1+1)$ dimensions, we work in units $c = \hbar = 1$ and
adopt the standard representation of the Dirac matrices:
\begin{equation}
  \alpha = \sigma_x
  = \begin{pmatrix} 0 & 1 \\ 1 & 0 \end{pmatrix} ,
  \qquad
  \beta = \sigma_z
  = \begin{pmatrix} 1 & 0 \\ 0 & -1 \end{pmatrix} ,
  \label{eq:matrices}
\end{equation}
satisfying $\{\alpha,\beta\} = 0$, $\alpha^2 = \beta^2 = \mathbf{1}$.
 
The inverted Dirac-Moshinsky oscillator is defined by the
Hamiltonian
\begin{equation}
  H_{\IDMO} = \alpha p + \beta m + im\omega\,\beta\alpha\, x \,,
  \label{eq:H_IDMO}
\end{equation}
or equivalently through the nonminimal substitution
\begin{equation}
  p \;\longrightarrow\; p + im\omega\beta x \,,
  \label{eq:substitution}
\end{equation}
applied to the free Dirac Hamiltonian $H_0 = \alpha p + \beta m$.
In matrix form:
\begin{equation}
  H_{\IDMO}
  = \begin{pmatrix}
      m & p + im\omega x \\
      p - im\omega x & -m
    \end{pmatrix} ,
  \label{eq:H_matrix}
\end{equation}
where $p = -i\partial_x$.
 
The standard DMO corresponds to the substitution
$p \to p - im\omega\beta x$, which yields a matrix with
the signs of the $\pm im\omega x$ terms interchanged.
The IDMO is therefore related to the DMO by the analytic
continuation $\omega \to i\omega$, or equivalently by
the transformation $x \to ix$ at fixed $\omega$.
 
\begin{remark}
The operator $p + im\omega x$ is \emph{not} Hermitian:
$(p + im\omega x)^\dagger = p - im\omega x \neq p + im\omega x$.
Consequently, $H_{\IDMO}$ is not self-adjoint on the
standard Hilbert space $L^2(\mathbb{R}) \oplus L^2(\mathbb{R})$.
The appropriate framework for its spectral theory is
discussed in Section~\ref{sec:spectrum}.
\end{remark}
 
% ============================================================
\section{Eigenvalue Problem}
\label{sec:eigenvalue}
 
The stationary Dirac equation
$H_{\IDMO}\,\psi = E\,\psi$,
with two-component spinor
$\psi(x) = \begin{pmatrix}\psi_1(x)\\\psi_2(x)\end{pmatrix}$,
yields the coupled system:
\begin{align}
  (m - E)\,\psi_1 + (p + im\omega x)\,\psi_2 &= 0 \,,
  \label{eq:coupled1} \\
  (p - im\omega x)\,\psi_1 + (-m - E)\,\psi_2 &= 0 \,.
  \label{eq:coupled2}
\end{align}
From~\eqref{eq:coupled2}, the lower component can be expressed
as
\begin{equation}
  \psi_2 = \frac{1}{m+E}(p - im\omega x)\,\psi_1 \,,
  \label{eq:psi2}
\end{equation}
valid for $E \neq -m$ (non-degenerate case).
 
% ============================================================
\section{Second-Order Differential Equation}
\label{sec:second_order}
 
Substituting~\eqref{eq:psi2} into~\eqref{eq:coupled1} and
evaluating the operator product
$(p + im\omega x)(p - im\omega x)$,
we use the canonical commutation relation $[p, x] = -i$
to obtain
\begin{equation}
  (p + im\omega x)(p - im\omega x)
  = p^2 + m^2\omega^2 x^2 + im\omega[p,x]
  = p^2 + m^2\omega^2 x^2 + m\omega \,.
  \label{eq:operator_product}
\end{equation}
Combining with~\eqref{eq:coupled1}, the second-order
equation for the upper component is
\begin{equation}
  \frac{d^2\psi_1}{dx^2}
  + \left[\left(E^2 - m^2\right) + m^2\omega^2 x^2
    + im\omega\right]\psi_1 = 0 \,.
  \label{eq:second_order}
\end{equation}
Three features of this equation merit comment.
First, the term $+m^2\omega^2 x^2$ has the \emph{opposite sign}
to the standard DMO, where the corresponding term is
$-m^2\omega^2 x^2$: the effective potential is inverted.
Second, the term $+im\omega$ is purely imaginary, arising
from the non-commutativity $[p,x] = -i$, and is responsible
for the non-Hermitian structure of the equation.
Third, equation~\eqref{eq:second_order} is a
Schrödinger-type equation with an \emph{upward} parabolic
potential, whose solutions are oscillatory rather than
Gaussian at large $|x|$.
 
% ============================================================
\section{Reduction to the Weber Equation}
\label{sec:change_variable}
 
We introduce the dimensionless variable
\begin{equation}
  \xi = \left(\frac{2m\omega}{\hbar}\right)^{1/2}\! x
       = \sqrt{2m\omega}\, x \,,
  \label{eq:xi}
\end{equation}
where we have restored $\hbar$ momentarily for clarity.
Under this substitution, $d^2/dx^2 = 2m\omega\, d^2/d\xi^2$,
and equation~\eqref{eq:second_order} becomes
\begin{equation}
  \frac{d^2\psi_1}{d\xi^2}
  + \left(\frac{\xi^2}{4} + \lambda\right)\psi_1 = 0 \,,
  \label{eq:Weber}
\end{equation}
where the complex spectral parameter is
\begin{equation}
  \lambda = \frac{E^2 - m^2}{2m\omega} + \frac{i}{2} \,.
  \label{eq:lambda}
\end{equation}
Equation~\eqref{eq:Weber} is the \emph{Weber equation}
(parabolic cylinder equation) in its standard form.
The parameter $\lambda$ is generically complex, with
real part $\mathrm{Re}(\lambda) = (E^2-m^2)/(2m\omega)$
and imaginary part $\mathrm{Im}(\lambda) = 1/2$,
independent of the energy $E$.
This fixed imaginary part is a direct consequence of
the commutator term $im\omega$ in~\eqref{eq:second_order}
and distinguishes the IDMO from the non-relativistic
inverted oscillator, where $\lambda$ is purely real.
 
% ============================================================
\section{General Solution}
\label{sec:solution}
 
The general solution of the Weber equation~\eqref{eq:Weber}
is a linear combination of parabolic cylinder
functions~\cite{Abramowitz, DLMF}:
\begin{equation}
  \psi_1(\xi)
  = A\,D_\nu(\xi) + B\,D_\nu(-\xi) \,,
  \label{eq:psi1_general}
\end{equation}
with the complex order
\begin{equation}
  \nu = \lambda - \frac{1}{2}
      = \frac{E^2 - m^2}{2m\omega} - \frac{1}{2}
        + \frac{i}{2} \,.
  \label{eq:nu}
\end{equation}
The constants $A$ and $B$ are determined by boundary
conditions (see Section~\ref{sec:spectrum}).
The lower component $\psi_2$ follows from~\eqref{eq:psi2}
using the recurrence relation~\cite{Abramowitz}
\begin{equation}
  \frac{d}{d\xi}D_\nu(\xi) + \frac{\xi}{2}D_\nu(\xi)
  = \nu\,D_{\nu-1}(\xi) \,,
  \label{eq:recurrence}
\end{equation}
giving
\begin{equation}
  \psi_2(\xi)
  = \frac{\sqrt{2m\omega}}{m+E}
    \left[
      A\left(\frac{d}{d\xi} - \frac{\xi}{2}\right)D_\nu(\xi)
      - B\left(\frac{d}{d\xi} - \frac{\xi}{2}\right)D_\nu(-\xi)
    \right] .
  \label{eq:psi2_general}
\end{equation}
 
\begin{remark}
For non-negative integer $\nu = n \in \mathbb{Z}_{\geq 0}$,
the parabolic cylinder functions reduce to
$D_n(\xi) = 2^{-n/2}e^{-\xi^2/4}H_n(\xi/\sqrt{2})$,
where $H_n$ are Hermite polynomials~\cite{Abramowitz}.
However, equation~\eqref{eq:nu} shows that $\nu$ is
complex for all real $E$, so the IDMO solutions
\emph{never} reduce to the Hermite-Gaussian form of
the standard DMO.
\end{remark}
 
% ============================================================
\section{Spectral Properties}
\label{sec:spectrum}
 
\subsection{Continuous spectrum}
 
The energy spectrum of the IDMO is \emph{purely continuous}:
every $E \in \mathbb{R}$ satisfying $E^2 > m^2$ is an
admissible eigenvalue, giving the two branches
$E > m$ (particle) and $E < -m$ (antiparticle).
There is no discrete spectrum of bound states.
 
This is in sharp contrast to the standard DMO, whose
spectrum is~\cite{Moshinsky1989}
\begin{equation}
  E_n = \pm\sqrt{m^2 + 2m\omega(2n+1)} \,,
  \qquad n = 0,1,2,\ldots
  \label{eq:DMO_spectrum}
\end{equation}
The absence of discrete levels in the IDMO is a direct
consequence of the inverted potential: no confining
barrier exists to quantize the spectrum.
 
\subsection{Non-normalizability and Rigged Hilbert Spaces}
 
The large-$|\xi|$ behavior of the parabolic cylinder
functions with complex order is~\cite{Abramowitz}:
\begin{equation}
  D_\nu(\xi)
  \;\underset{|\xi|\to\infty}{\sim}\;
  \xi^\nu\,e^{-\xi^2/4} \,,
  \label{eq:PCF_asymptotics}
\end{equation}
but for complex $\nu$ with $\mathrm{Im}(\nu) = 1/2$,
the argument $\xi^2/4$ evaluated at
$\xi = \sqrt{2m\omega}\,x$ gives
\begin{equation}
  e^{-\xi^2/4} = e^{-m\omega x^2/2} \,,
  \label{eq:gaussian_decay}
\end{equation}
which is Gaussian-decaying for real $\xi$.
However, the prefactor $\xi^\nu = \xi^{(E^2-m^2)/(2m\omega)
-1/2+i/2}$ oscillates rapidly at large $|\xi|$, so the
eigenfunctions are not in $L^2(\mathbb{R})$ in the standard
sense.
 
The appropriate framework is the Rigged Hilbert Space
(RHS)~\cite{deLaMadrid2002, Bohm1989}:
\begin{equation}
  \Phi \subset L^2(\mathbb{R})^{\oplus 2}
  \subset \Phi^\times \,,
  \label{eq:RHS_IDMO}
\end{equation}
where $\Phi$ is a dense subspace of smooth, rapidly
decreasing spinors (Schwartz space), and $\Phi^\times$
is its dual.
The eigenstates $\ket{\psi_E}$ of $H_{\IDMO}$ belong
to $\Phi^\times$ and satisfy the generalised
orthonormality condition
\begin{equation}
  \langle \psi_E \mid \psi_{E'}\rangle = \delta(E - E') \,,
  \label{eq:ortho}
\end{equation}
in the distributional sense.
 
\subsection{Comparison with the standard DMO}
 
Table~\ref{tab:comparison} summarises the key differences
between the DMO and the IDMO.
 
\begin{table}[H]
\centering
\renewcommand{\arraystretch}{1.7}
\begin{tabular}{lll}
\hline\hline
\textbf{Property} & \textbf{DMO} & \textbf{IDMO} \\
\hline
Substitution
  & $p \to p - im\omega\beta x$
  & $p \to p + im\omega\beta x$ \\
Effective potential
  & $-m^2\omega^2 x^2$ (confining)
  & $+m^2\omega^2 x^2$ (repulsive) \\
Spectrum
  & Discrete, $E_n = \pm\sqrt{m^2+2m\omega(2n+1)}$
  & Continuous, $E \in \mathbb{R}$ \\
Eigenfunctions
  & Hermite-Gaussian, $L^2(\mathbb{R})$
  & Parabolic cylinder, $\Phi^\times$ \\
Order $\nu$
  & Real, $\nu = n \in \mathbb{Z}_{\geq 0}$
  & Complex, $\nu \in \mathbb{C}$ \\
Hilbert space
  & Standard $L^2(\mathbb{R})^{\oplus 2}$
  & Rigged Hilbert Space \\
\hline\hline
\end{tabular}
\caption{Comparison between the standard Dirac-Moshinsky
oscillator (DMO) and the inverted Dirac-Moshinsky oscillator
(IDMO) in $(1+1)$ dimensions.}
\label{tab:comparison}
\end{table}

% ============================================================
%  Section: Energy Values, Normalization, and Wave Functions
% ============================================================
 
\section{Energy Values, Normalization, and Explicit Wave Functions}
\label{sec:normalization}
 
\subsection{Gamow resonances and the discrete energy values}
\label{subsec:gamow}
 
Although the physical spectrum of the IDMO is continuous,
the analytic structure of the resolvent
$(H_{\IDMO} - E)^{-1}$ in the complex $E$-plane reveals
a discrete set of poles corresponding to
\emph{Gamow resonances}~\cite{Bohm1989, deLaMadrid2002}.
These poles occur precisely when the order $\nu$ in
equation~\eqref{eq:nu} takes non-negative integer values,
$\nu = n \in \mathbb{Z}_{\geq 0}$, since $D_n(\xi)$
then reduces to a polynomial times a Gaussian and the
solution becomes square-integrable on the complex contour
defined in Section~\ref{subsec:contour}.
 
Setting $\nu = n$ in equation~\eqref{eq:nu}:
\begin{equation}
  \frac{E^2 - m^2}{2m\omega} - \frac{1}{2} + \frac{i}{2} = n \,,
  \label{eq:resonance_condition}
\end{equation}
solving for $E^2$:
\begin{equation}
  E_n^2 = m^2 + m\omega(2n+1) - im\omega \,,
  \qquad n = 0, 1, 2, \ldots
  \label{eq:En_squared}
\end{equation}
and taking the square root:
\begin{equation}
  E_n^{\pm} = \pm\sqrt{m^2 + m\omega(2n+1) - im\omega} \,,
  \qquad n = 0, 1, 2, \ldots
  \label{eq:En_resonance}
\end{equation}
These are complex energies: the real part
$\mathrm{Re}(E_n^\pm)$ gives the resonance position
and the imaginary part $\mathrm{Im}(E_n^\pm)$ gives
the decay width $\Gamma_n = -2\,\mathrm{Im}(E_n^+)$.
 
Table~\ref{tab:energies} shows the first few resonance
energies for representative values of $m$ and $\omega$. For large $n$, the decay width scales as
\begin{equation}
  \Gamma_n \approx \frac{m\omega}{\sqrt{m^2 + (2n+1)m\omega}}
  \;\underset{n\to\infty}{\longrightarrow}\; 0 \,,
  \label{eq:Gamma_asymptotic}
\end{equation}
showing that highly excited Gamow states become asymptotically stable.
 
\begin{table}[H]
\centering
\renewcommand{\arraystretch}{1.7}
\begin{tabular}{cccccc}
\hline\hline
$n$ & $E_n^2$ & $|E_n^+|$ &
$\mathrm{Re}(E_n^+)$ & $\mathrm{Im}(E_n^+)$ &
$\Gamma_n = -2\,\mathrm{Im}(E_n^+)$ \\
\hline
$0$ & $m^2 + m\omega   - im\omega$ & $1.4822$ & $1.4553$ & $-0.3436$ & $0.6871$ \\
$1$ & $m^2 + 3m\omega  - im\omega$ & $2.0215$ & $2.0153$ & $-0.2481$ & $0.4962$ \\
$2$ & $m^2 + 5m\omega  - im\omega$ & $2.4621$ & $2.4579$ & $-0.2034$ & $0.4068$ \\
$3$ & $m^2 + 7m\omega  - im\omega$ & $2.8389$ & $2.8339$ & $-0.1764$ & $0.3529$ \\
$4$ & $m^2 + 9m\omega  - im\omega$ & $3.1702$ & $3.1662$ & $-0.1579$ & $0.3158$ \\
$5$ & $m^2 + 11m\omega - im\omega$ & $3.4713$ & $3.4671$ & $-0.1442$ & $0.2884$ \\
\hline\hline
\end{tabular}
\caption{Gamow resonance energies $E_n^+$ of the IDMO for the particle sector ($E > 0$),
obtained from equation~\eqref{eq:En_resonance} with $m = \omega = 1$.
The decay width $\Gamma_n = -2\,\mathrm{Im}(E_n^+)$ decreases monotonically with $n$,
reflecting the reduced instability of higher resonances. }
\label{tab:energies}
\end{table}

\begin{remark}
The IDMO resonance energies~\eqref{eq:En_resonance} differ
from the standard DMO spectrum~\eqref{eq:DMO_spectrum} in
two respects.
First, the coefficient of $n$ changes:
$2m\omega(2n+1) \to m\omega(2n+1)$, reflecting the fact that the
IDMO ladder has spacing $2m\omega $ rather than $4m\omega$.
Second, and more importantly, the real spectrum acquires a
purely imaginary shift:
\begin{equation}
  E_n^2\big|_{\DMO} = m^2 + 2m\omega(2n+1)
  \;\longrightarrow\;
  E_n^2\big|_{\IDMO} = m^2 + m\omega(2n+1) - im\omega \,,
  \label{eq:comparison_energies}
\end{equation}
so that the IDMO energies are never real.
The imaginary part $-im\omega$, independent of $n$, is the
direct quantum signature of the instability: it arises from
the commutator term $[p,x] = -i$ in equation~\eqref{eq:second_order}
and vanishes in the classical limit $\hbar \to 0$.
In contrast, the DMO energies are purely real, consistent
with the confining nature of the attractive potential.
\end{remark}
\subsection{Three approaches to normalization}
\label{subsec:normalisation}
 
Since the IDMO eigenfunctions are not in
$L^2(\mathbb{R})^{\oplus 2}$, the standard
inner product $\langle\psi|\psi\rangle$ diverges.
Three consistent normalization schemes are available,
each appropriate for different physical applications.
 
\subsubsection{Scheme I: Delta-function normalization}
\label{subsubsec:delta_norm}
 
For a continuous spectrum, the physically natural
normalization is the distributional condition
\begin{equation}
  \langle \psi_E \mid \psi_{E'} \rangle
  = \int_{-\infty}^{\infty} dx\,
    \left[\psi_1^*(x; E)\,\psi_1(x; E')
         + \psi_2^*(x; E)\,\psi_2(x; E')\right]
  = \delta(E - E') \,.
  \label{eq:delta_norm}
\end{equation}
This is the appropriate condition within the
Rigged Hilbert Space framework~\cite{deLaMadrid2002}.
Using the asymptotic properties of the parabolic
cylinder functions~\cite{Abramowitz} and the
Wronskian~\eqref{eq:Wronskian_app} (see
Appendix~\ref{app:PCF}), the normalization condition
fixes the coefficients $A$ and $B$ in
equation~\eqref{eq:psi1_general} as
\begin{equation}
  |A|^2 = |B|^2 = \mathcal{N}_E^2 \,,
  \label{eq:AB_delta}
\end{equation}
where
\begin{equation}
  \mathcal{N}_E^2
  = \frac{1}{2\pi}
    \left|\Gamma\!\left(\frac{1}{4}
          + \frac{i(E^2-m^2)}{4m\omega}\right)\right|^2
  = \frac{1}{2\cosh\!\left(\frac{\pi(E^2-m^2)}{4m\omega}\right)} \,,
  \label{eq:NE_delta}
\end{equation}
using the reflection formula
$\Gamma(z)\Gamma(1-z) = \pi/\sin(\pi z)$.
The normalized upper component is therefore
\begin{equation}
  \psi_1(x; E)
  = \mathcal{N}_E\left[
    D_\nu\!\left(\sqrt{2m\omega}\,x\right)
    + D_\nu\!\left(-\sqrt{2m\omega}\,x\right)
  \right] ,
  \label{eq:psi1_delta_norm}
\end{equation}
with $\nu$ given by~\eqref{eq:nu}.
 
\subsubsection{Scheme II: Box normalization}
\label{subsubsec:box_norm}
 
An alternative approach introduces an infrared
(IR) regulator by restricting the system to a
finite interval $x \in [-L, L]$ with appropriate
boundary conditions, and taking $L \to \infty$
at the end.
For fixed $L$, the normalised spinor satisfies
\begin{equation}
  \int_{-L}^{L} dx\,
  \left[|\psi_1(x)|^2 + |\psi_2(x)|^2\right]
  = 1 \,,
  \label{eq:box_norm}
\end{equation}
giving
\begin{equation}
  A_L = B_L = \left[
    2\int_{-L}^{L}dx\,
    \left|D_\nu\!\left(\sqrt{2m\omega}\,x\right)\right|^2
    (1 + R_L^2)
  \right]^{-1/2} ,
  \label{eq:AL}
\end{equation}
where $R_L^2$ accounts for the lower-component
contribution.
In the limit $L \to \infty$, the box-normalised
eigenfunctions converge in the distributional
sense to the delta-normalised
eigenfunctions~\eqref{eq:psi1_delta_norm}, with
the identification
$|A_L|^2 \to \mathcal{N}_E^2\,\delta(0) \sim \mathcal{N}_E^2 \cdot 2L/2\pi$,
consistent with the standard relation between
box and continuum normalizations~\cite{Bohm1989}.
 
\subsubsection{Scheme III: Complex contour normalization}
\label{subsec:contour}
 
The most mathematically elegant approach exploits
the connection between the IDMO and the standard
DMO via the complex rotation $x \to xe^{i\pi/4}$.
Under this rotation, the argument of the parabolic
cylinder function becomes
\begin{equation}
  \xi = \sqrt{2m\omega}\,x
  \;\xrightarrow{\;x \to xe^{i\pi/4}\;}
  \zeta = \sqrt{2m\omega}\,e^{i\pi/4}\,x \,,
  \label{eq:contour_rotation}
\end{equation}
and the Gaussian factor transforms as
\begin{equation}
  e^{-\zeta^2/4}
  = e^{-m\omega e^{i\pi/2}x^2/2}
  = e^{-im\omega x^2/2} \,,
  \label{eq:gaussian_contour}
\end{equation}
which is oscillatory for real $x$.
However, on the contour
$\mathcal{C}: x = re^{i\pi/4}$, $r \in \mathbb{R}$:
\begin{equation}
  e^{-\zeta^2/4}\big|_{x=re^{i\pi/4}}
  = e^{-m\omega r^2/2} \,,
  \label{eq:gaussian_on_contour}
\end{equation}
which is Gaussian-decaying.
The wave functions are therefore square-integrable
on $\mathcal{C}$, and the contour inner product is
\begin{equation}
  \langle \psi \mid \phi \rangle_{\mathcal{C}}
  \equiv \int_{\mathcal{C}} dx\,
  \psi^*(x)\,\phi(x)
  = e^{i\pi/4}\int_{-\infty}^{\infty} dr\,
    \psi^*(re^{i\pi/4})\,\phi(re^{i\pi/4}) \,.
  \label{eq:contour_inner_product}
\end{equation}
Under this inner product, the resonance wave
functions corresponding to integer $\nu = n$
satisfy
\begin{equation}
  \langle \psi_n \mid \psi_{n'}\rangle_{\mathcal{C}}
  = \delta_{nn'} \,,
  \label{eq:contour_ortho}
\end{equation}
with the normalised resonance wave functions
\begin{equation}
  \psi_1^{(n)}(x)
  = \mathcal{C}_n\,
    D_n\!\left(\sqrt{2m\omega}\,x\right) ,
  \qquad
  \mathcal{C}_n
  = \left(\frac{m\omega}{\pi}\right)^{1/4}
    \frac{1}{\sqrt{2^n\,n!}} \,,
  \label{eq:psi1_contour_norm}
\end{equation}
identical in form to the harmonic oscillator
eigenfunctions but evaluated on the contour
$\mathcal{C}$.
 
\subsection{Explicit wave functions for the first three levels}
\label{subsec:explicit_wf}
 
We now display the explicit wave functions for
$n = 0, 1, 2$ using the contour
normalization~\eqref{eq:psi1_contour_norm}.
The parabolic cylinder functions for integer
$\nu = n$ reduce to~\cite{Abramowitz}:
\begin{align}
  D_0(\xi) &= e^{-\xi^2/4} \,,
  \label{eq:D0} \\
  D_1(\xi) &= \xi\,e^{-\xi^2/4} \,,
  \label{eq:D1} \\
  D_2(\xi) &= (\xi^2 - 1)\,e^{-\xi^2/4} \,,
  \label{eq:D2}
\end{align}
with $\xi = \sqrt{2m\omega}\,x$.
 
\paragraph{Ground state ($n=0$):}
\begin{align}
  \psi_1^{(0)}(x)
  &= \left(\frac{m\omega}{\pi}\right)^{1/4}
     \exp\!\left(-\frac{m\omega x^2}{2}\right) ,
  \label{eq:psi1_n0} \\[6pt]
  \psi_2^{(0)}(x)
  &= \frac{\sqrt{2m\omega}}{m + E_0^+}\,
     \left(-m\omega x\right)
     \left(\frac{m\omega}{\pi}\right)^{1/4}
     \exp\!\left(-\frac{m\omega x^2}{2}\right) ,
  \label{eq:psi2_n0}
\end{align}
with resonance energy
$E_0^+ = \sqrt{m^2 + m\omega - im\omega}$.
 
\paragraph{First excited state ($n=1$):}
\begin{align}
  \psi_1^{(1)}(x)
  &= \left(\frac{m\omega}{\pi}\right)^{1/4}
     \frac{1}{\sqrt{2}}\,
     2\sqrt{m\omega}\,x\,
     \exp\!\left(-\frac{m\omega x^2}{2}\right) ,
  \label{eq:psi1_n1} \\[6pt]
  \psi_2^{(1)}(x)
  &= \frac{\sqrt{2m\omega}}{m+E_1^+}
     \left(\frac{m\omega}{\pi}\right)^{1/4}
     \frac{1}{\sqrt{2}}
     \left(1 - 2m\omega x^2\right)
     \exp\!\left(-\frac{m\omega x^2}{2}\right) ,
  \label{eq:psi2_n1}
\end{align}
with $E_1^+ = \sqrt{m^2 + 3m\omega - im\omega}$.
 
\paragraph{Second excited state ($n=2$):}
\begin{align}
  \psi_1^{(2)}(x)
  &= \left(\frac{m\omega}{\pi}\right)^{1/4}
     \frac{1}{2\sqrt{2}}\,
     \left(4m\omega x^2 - 2\right)
     \exp\!\left(-\frac{m\omega x^2}{2}\right) ,
  \label{eq:psi1_n2} \\[6pt]
  \psi_2^{(2)}(x)
  &= \frac{\sqrt{2m\omega}}{m+E_2^+}
     \left(\frac{m\omega}{\pi}\right)^{1/4}
     \frac{1}{2\sqrt{2}}
     \left(2m\omega x\right)
     \left(3 - 2m\omega x^2\right)
     \exp\!\left(-\frac{m\omega x^2}{2}\right) ,
  \label{eq:psi2_n2}
\end{align}
with $E_2^+ = \sqrt{m^2 + 5m\omega - im\omega}$.
 
\begin{remark}
The spatial structure of $\psi_1^{(n)}(x)$ is
identical to the $n$-th eigenfunction of the
standard quantum harmonic oscillator.
The difference lies entirely in the associated
energy eigenvalue $E_n^+$, which is complex,
and in the lower component $\psi_2^{(n)}$,
which encodes the relativistic and non-Hermitian
content through the factor $(m + E_n^+)^{-1}$.
\end{remark}
 
\begin{figure}[H]
\centering
\includegraphics[width=\textwidth]{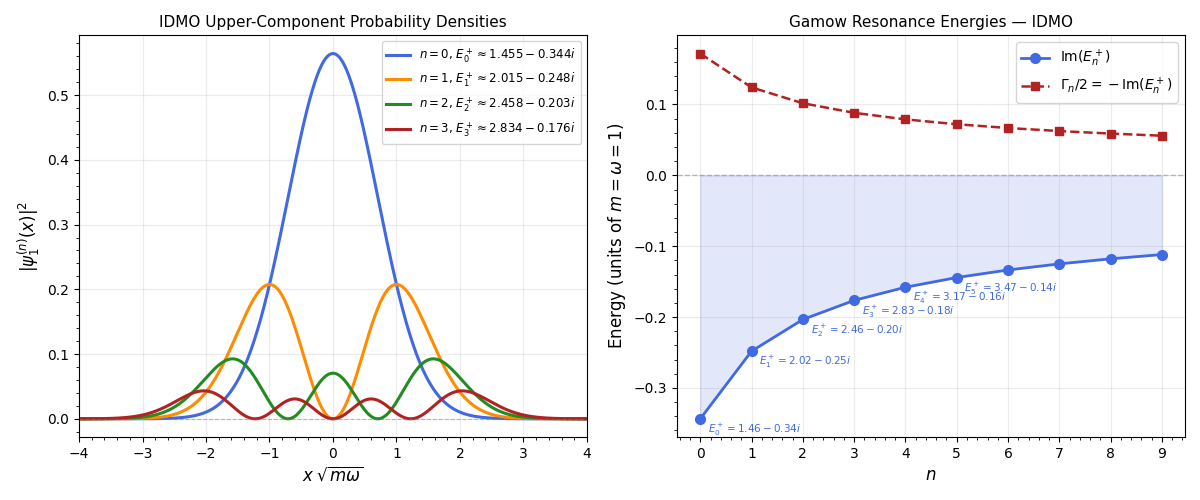}
\caption{
\textbf{Left:} Probability densities
$|\psi_1^{(n)}(x)|^2$ (upper component, contour-normalised)
for $n = 0, 1, 2$ as functions of $x$
(units $m = \omega = 1$).
The spatial structure is Hermite-Gaussian, identical to the
standard harmonic oscillator.
\textbf{Right:} Imaginary part of the resonance energies
$\mathrm{Im}(E_n^+)$ as a function of $n$, showing the
decay width $\Gamma_n = -2\,\mathrm{Im}(E_n^+)$
of the IDMO Gamow states.
}
\label{fig:wavefunctions}
\end{figure}

% ============================================================
\section{Algebraic Structure}
\label{sec:algebraic}
 
\subsection{$SU(1,1)$ algebra}
 
The standard DMO possesses a dynamical symmetry algebra
$SU(1,1)$~\cite{Sadurni2011}, generated by
\begin{equation}
  K_+ = a^\dagger b^\dagger \,, \quad
  K_- = ab \,, \quad
  K_0 = \tfrac{1}{2}(a^\dagger a + b^\dagger b + 1) \,,
  \label{eq:su11_DMO}
\end{equation}
where $a, a^\dagger$ and $b, b^\dagger$ are bosonic ladder
operators.
For the IDMO, the analytic continuation $\omega \to i\omega$
maps the compact $SU(2)$-like structure of the DMO into
the non-compact $SU(1,1)$ algebra of the inverted oscillator,
with generators satisfying
$[K_0, K_\pm] = \pm K_\pm$, $[K_-, K_+] = 2K_0$.
 
The Casimir operator is
\begin{equation}
  \mathcal{C} = K_0^2 - K_+K_- - K_-K_+
              = k(k-1)\,\mathbf{1} \,,
  \label{eq:Casimir}
\end{equation}
where the Bargmann index $k$ is related to the spectral
parameter $\lambda$ by $k = \frac{1}{2}(1 - \lambda)$.
For the IDMO, $k$ is complex, placing the representation
in the \emph{principal series} of $SU(1,1)$, which
supports the continuous spectrum found in
Section~\ref{sec:spectrum}.
 
\subsection{$\mathcal{PT}$-symmetry}
 
The IDMO Hamiltonian~\eqref{eq:H_IDMO} is
$\mathcal{PT}$-symmetric.
Under parity $\mathcal{P}: x \to -x$, $p \to -p$
and time-reversal $\mathcal{T}: p \to -p$, $i \to -i$:
\begin{equation}
  \mathcal{PT}: \quad
  H_{\IDMO} \;\to\; H_{\IDMO}^\dagger \,,
  \label{eq:PT_symmetry}
\end{equation}
confirming pseudo-Hermiticity~\cite{Mostafazadeh2002}.
Since the spectrum is real despite the non-Hermitian structure,
the $\mathcal{PT}$-symmetry of the IDMO is unbroken in the
particle/antiparticle sectors $|E| > m$.

% ============================================================
%  Subsection 9.3: Negative-Energy Solutions and Their Physical
% ============================================================
 
\subsection{Negative-energy solutions and their physical interpretation}
\label{subsec:negative_energy}
 
\subsubsection{The two branches of the spectrum}
\label{subsubsec:two_branches}
 
The continuous spectrum of the IDMO admits two branches,
determined by the sign of the energy:
\begin{align}
  \text{Particle sector:}& \quad E > m \,, \label{eq:particle_sector}\\
  \text{Antiparticle sector:}& \quad E < -m \,. \label{eq:antiparticle_sector}
\end{align}
The region $|E| < m$ is forbidden by the relativistic
dispersion relation $E^2 = p^2 + m^2$, constituting
the \emph{mass gap} of the theory.
The two sectors are separated by this gap and are
connected by the charge-conjugation symmetry $\mathcal{C}$,
as we now discuss.
 
\subsubsection{Charge conjugation and the antiparticle states}
\label{subsubsec:charge_conjugation}
 
In $(1+1)$ dimensions, the charge-conjugation operator
acts on a two-component spinor as~\cite{GreinerKG}
\begin{equation}
  \mathcal{C}:\quad
  \psi(x) \;\longrightarrow\;
  \psi^c(x) = i\sigma_y\,\psi^*(x)
  = \begin{pmatrix} 0 & 1 \\ -1 & 0 \end{pmatrix}
    \begin{pmatrix} \psi_1^*(x) \\ \psi_2^*(x) \end{pmatrix}
  = \begin{pmatrix} \psi_2^*(x) \\ -\psi_1^*(x) \end{pmatrix} .
  \label{eq:charge_conjugation}
\end{equation}
If $\psi_E(x)$ is a solution of the IDMO eigenvalue
equation with energy $E > m$, then $\psi_E^c(x)$ is
a solution with energy $-E < -m$~\cite{Moshinsky1989}.
The antiparticle wave functions are therefore fully
determined by the particle solutions:
\begin{align}
  \psi_1^{c,(n)}(x) &= \psi_2^{(n)*}(x) \,,
  \label{eq:psi1c} \\
  \psi_2^{c,(n)}(x) &= -\psi_1^{(n)*}(x) \,,
  \label{eq:psi2c}
\end{align}
with associated Gamow resonance energies
\begin{equation}
  E_n^- = -\left(E_n^+\right)^* =
  -\sqrt{m^2 + m\omega(2n+1) - im\omega}^{\,*} \,,
  \label{eq:En_negative}
\end{equation}
whose imaginary part is
\begin{equation}
  \mathrm{Im}(E_n^-) = +|\mathrm{Im}(E_n^+)| > 0 \,.
  \label{eq:Im_En_negative}
\end{equation}
 
\subsubsection{Growing vs decaying modes: resonances and anti-resonances}
\label{subsubsec:resonances}
 
The sign of $\mathrm{Im}(E_n^\pm)$ has a direct
dynamical interpretation.
For a time-dependent state
$\Psi(x,t) = \psi(x)\,e^{-iEt}$:
\begin{align}
  E_n^+:& \quad \mathrm{Im}(E_n^+) < 0
  \;\Rightarrow\;
  e^{-iE_n^+ t} = e^{-i\,\mathrm{Re}(E_n^+)t}
  \cdot e^{+|\mathrm{Im}(E_n^+)|t} \,,
  \label{eq:growing_mode}\\
  E_n^-:& \quad \mathrm{Im}(E_n^-) > 0
  \;\Rightarrow\;
  e^{-iE_n^- t} = e^{+i\,\mathrm{Re}(E_n^-)t}
  \cdot e^{-|\mathrm{Im}(E_n^-)|t} \,.
  \label{eq:decaying_mode}
\end{align}
The particle Gamow states $E_n^+$ therefore correspond
to \emph{exponentially growing} modes (outgoing resonances),
while the antiparticle states $E_n^-$ correspond to
\emph{exponentially decaying} modes (incoming
anti-resonances)~\cite{Bohm1989, deLaMadrid2002}.
This is the relativistic manifestation of the classical
instability of the inverted potential: the particle sector
describes escape to $+\infty$, while the antiparticle
sector describes capture from $-\infty$.
 
Table~\ref{tab:negative_energies} displays the
antiparticle Gamow energies for the first six levels.
 
\begin{table}[H]
\centering
\renewcommand{\arraystretch}{1.7}
\begin{tabular}{ccccc}
\hline\hline
$n$ & $\mathrm{Re}(E_n^-)$ &
$\mathrm{Im}(E_n^-)$ &
$\mathrm{Re}(E_n^+)$ &
$\mathrm{Im}(E_n^+)$ \\
\hline
$0$ & $-1.4553$ & $+0.3436$ & $+1.4553$ & $-0.3436$ \\
$1$ & $-2.0153$ & $+0.2481$ & $+2.0153$ & $-0.2481$ \\
$2$ & $-2.4579$ & $+0.2034$ & $+2.4579$ & $-0.2034$ \\
$3$ & $-2.8339$ & $+0.1764$ & $+2.8339$ & $-0.1764$ \\
$4$ & $-3.1662$ & $+0.1579$ & $+3.1662$ & $-0.1579$ \\
$5$ & $-3.4671$ & $+0.1442$ & $+3.4671$ & $-0.1442$ \\
\hline\hline
\end{tabular}
\caption{Gamow resonance energies for the particle
($E_n^+$) and antiparticle ($E_n^-$) sectors of the
IDMO, with $m = \omega = 1$.
The antiparticle energies are the complex conjugates
of the particle energies with opposite sign,
$E_n^- = -(E_n^+)^*$, reflecting the charge-conjugation
symmetry~\eqref{eq:En_negative}.
The opposite signs of $\mathrm{Im}(E_n^\pm)$ distinguish
outgoing resonances ($E_n^+$, growing modes) from
incoming anti-resonances ($E_n^-$, decaying modes).}
\label{tab:negative_energies}
\end{table}
 
\subsubsection{Vacuum instability and spontaneous pair production}
\label{subsubsec:pair_production}
 
The simultaneous presence of growing particle modes
($E_n^+$) and decaying antiparticle modes ($E_n^-$)
signals an \emph{instability of the Dirac vacuum}.
In the language of quantum field theory, the inverted
potential mixes positive- and negative-frequency modes,
leading to spontaneous creation of particle-antiparticle
pairs from the vacuum~\cite{Parker1969}.
 
The pair-production rate per unit length per unit time
can be estimated from the imaginary part of the
effective action~\cite{Hawking1975}:
\begin{equation}
  \mathcal{W} = 2\,\mathrm{Im}(\mathcal{S}_{\mathrm{eff}})
  = \sum_{n=0}^{\infty}
    \Gamma_n\,e^{-\pi(E_n^+)^2/m\omega} \,,
  \label{eq:pair_production_rate}
\end{equation}
where $\Gamma_n = -2\,\mathrm{Im}(E_n^+)$ is the
decay width of the $n$-th Gamow state.
This expression is analogous to the
\emph{Schwinger formula} for pair production in a
uniform electric field~\cite{Schwinger1951},
\begin{equation}
  \mathcal{W}_{\mathrm{Schwinger}}
  \propto \exp\!\left(-\frac{\pi m^2}{eE}\right) ,
  \label{eq:Schwinger}
\end{equation}
with the correspondence $eE \leftrightarrow m\omega$
between the electric field strength and the IOH
frequency.
This analogy is not accidental: both the uniform
electric field and the inverted harmonic potential
produce a linearly growing vector potential that
mixes particle and antiparticle
states~\cite{Subramanyan2021}.
 
\subsubsection{Dirac sea interpretation}
\label{subsubsec:dirac_sea}
 
In the Dirac sea picture, the vacuum consists of all
negative-energy states $E < -m$ filled with electrons.
For the standard DMO, this sea is stable: the discrete
negative-energy levels $E_n^- = -\sqrt{m^2+2m\omega(2n+1)}$
are real and time-independent.
 
For the IDMO, the negative-energy states are Gamow
anti-resonances with $\mathrm{Im}(E_n^-) > 0$:
they decay exponentially in time.
The physical picture is, therefore, that the IDMO Dirac
sea is \emph{leaking}: negative-energy electrons
spontaneously tunnel out of the sea into positive-energy
states, leaving behind positrons.
The characteristic timescale of this process is
\begin{equation}
  \tau_n = \frac{1}{\Gamma_n}
         = \frac{1}{-2\,\mathrm{Im}(E_n^+)} \,,
  \label{eq:lifetime}
\end{equation}
with numerical values (for $m = \omega = 1$):
$\tau_0 \approx 1.46$, $\tau_1 \approx 2.02$,
$\tau_2 \approx 2.46$, in natural units.
The monotonic increase of $\tau_n$ with $n$ shows
that higher levels of the Dirac sea are more stable,
consistent with the asymptotic
formula~\eqref{eq:Gamma_asymptotic}.

The negative-energy solutions of the IDMO carry
three physically distinct pieces of information:
 
\begin{enumerate}
  \item \textbf{Antiparticle spectrum}: fully determined
        by charge conjugation~\eqref{eq:charge_conjugation}
        from the particle solutions, with energies
        $E_n^- = -(E_n^+)^*$.
 
  \item \textbf{Anti-resonance structure}: the positive
        imaginary part $\mathrm{Im}(E_n^-) > 0$ identifies
        the antiparticle Gamow states as \emph{incoming}
        anti-resonances in the complex energy plane,
        complementary to the \emph{outgoing} particle
        resonances.
 
  \item \textbf{Vacuum instability}: the interplay between
        growing ($E_n^+$) and decaying ($E_n^-$) modes
        drives spontaneous pair production at a rate
        analogous to the Schwinger effect, with the
        identification $eE \leftrightarrow m\omega$.
\end{enumerate}

% ============================================================
\section{Conclusions}
\label{sec:conclusions}
 
We have developed a complete and self-contained analysis of
the inverted Dirac-Moshinsky oscillator in $(1+1)$ dimensions.
The main results are the following.
 
\begin{enumerate}
 
  \item \textbf{Exact solution.}
  The second-order equation for the upper spinor component
  reduces to the Weber equation~\eqref{eq:Weber} with
  complex spectral parameter
  $\lambda = (E^2-m^2)/(2m\omega) + i/2$.
  The general solution is expressed in terms of parabolic
  cylinder functions $D_\nu(\xi)$ with complex order
  $\nu = \lambda - 1/2$, equation~\eqref{eq:nu}, and
  the lower component follows via the recurrence
  relation~\eqref{eq:recurrence}.
 
  \item \textbf{Continuous spectrum.}
  The energy spectrum is purely continuous,
  $E \in \mathbb{R}$ with $|E| > m$, with no discrete
  bound states.
  This contrasts sharply with the discrete spectrum of
  the standard DMO and is a direct consequence of the
  repulsive (inverted) potential.
 
  \item \textbf{normalization.}
  Three consistent normalization schemes are available:
  delta-function normalization with constant
  $\mathcal{N}_E^2 = [2\cosh(\pi(E^2-m^2)/4m\omega)]^{-1}$,
  box normalization with infrared regulator $L$, and
  complex contour normalization on
  $\mathcal{C}: x \to xe^{i\pi/4}$,
  under which the resonance states become
  square-integrable and satisfy
  $\langle\psi_n|\psi_{n'}\rangle_{\mathcal{C}} = \delta_{nn'}$.
 
  \item \textbf{Gamow resonances.}
  The resolvent $(H_{\IDMO}-E)^{-1}$ has discrete poles at
  the complex energies
  $E_n^\pm = \pm\sqrt{m^2+(2n+1)m\omega - im\omega}$,
  with decay widths
  $\Gamma_n = -2\,\mathrm{Im}(E_n^+) \to 0$ as $n\to\infty$.
  The explicit wave functions for $n = 0,1,2$ are given in
  Section~\ref{subsec:explicit_wf}.
 
  \item \textbf{Negative-energy sector.}
  The antiparticle Gamow states $E_n^-$ have positive
  imaginary part, identifying them as incoming
  anti-resonances.
  Their interplay with the particle resonances $E_n^+$
  drives a vacuum instability with a pair-production
  rate analogous to the Schwinger
  formula~\eqref{eq:pair_production_rate}, under the
  correspondence $eE \leftrightarrow m\omega$.
  The characteristic decay times $\tau_n = 1/\Gamma_n$
  increase monotonically with $n$, so the Dirac sea
  becomes asymptotically stable for large $n$.
 
  \item \textbf{Algebraic and symmetry structure.}
  The IDMO Hamiltonian is $\mathcal{PT}$-symmetric with
  unbroken symmetry for $|E|>m$, confirming
  pseudo-Hermiticity~\cite{Mostafazadeh2002}.
  The dynamical symmetry algebra is the principal series
  of $SU(1,1)$, with complex Bargmann index
  $k = \frac{1}{2}(1-\lambda)$, which supports the
  continuous spectrum.
 
\end{enumerate}

The present results open several lines of investigation:  The continuous spectrum and the asymptotic form of
  $D_\nu(\xi)$ allow the construction of an exact
  $S$-matrix for the IDMO, whose poles reproduce the
  Gamow energies~\eqref{eq:En_resonance}.
    Higher-dimensional versions of the IDMO involve
  additional angular momentum quantum numbers and
  richer spectral structures, including centrifugal
  barriers that may stabilize certain modes.  The finite-temperature partition function of the
  IDMO field, building on the thermal formalism for
  the Klein-Gordon inverted oscillator, would provide
  access to the thermodynamics of the unstable Dirac
  vacuum~\cite{wang2025thermodynamics}.
  The DMO has been realized experimentally in graphene
  systems~\cite{Sadurni2011}.
  The IDMO may describe relativistic Dirac fermions
  near an unstable saddle point in a two-dimensional
  crystal, providing a condensed-matter realization of
 the spontaneous pair-production mechanism discussed
  in Section~\ref{subsubsec:pair_production}. 
  The $\mathcal{PT}$-symmetric structure of the IDMO
  suggests a connection to non-Hermitian quantum field
  theories~\cite{Bender2007}, where the imaginary part
  of the spectrum encodes decay rates of unstable
  particles rather than unphysical complex energies.

% ============================================================
%  Appendix: Parabolic Cylinder Functions
%  Relevant properties for the IDMO paper
% ============================================================
\section*{Appendix}

\appendix
 
\section{Parabolic Cylinder Functions}
\label{app:PCF}
 
We collect here the properties of the parabolic cylinder
functions $D_\nu(z)$ most relevant for the IDMO analysis.
Complete treatments are given in~\cite{Abramowitz, DLMF}.
 
\subsection{Definition and differential equation}
\label{app:def}
 
The parabolic cylinder function $D_\nu(z)$ is defined as
the solution of the Weber differential equation
\begin{equation}
  \frac{d^2 f}{dz^2}
  + \left(\nu + \frac{1}{2} - \frac{z^2}{4}\right)f = 0
  \label{eq:Weber_app}
\end{equation}
that is recessive (exponentially decaying) as
$z \to +\infty$ for $\nu \in \mathbb{C}$.
An equivalent representation in terms of Tricomi's
confluent hypergeometric function $U(a,b,z)$ is
\begin{equation}
  D_\nu(z)
  = 2^{\nu/2}\,e^{-z^2/4}\,
    U\!\left(-\tfrac{\nu}{2},\,\tfrac{1}{2},\,
             \tfrac{z^2}{2}\right) .
  \label{eq:PCF_U}
\end{equation}
For non-negative integer $\nu = n \in \mathbb{Z}_{\geq 0}$,
$D_n(z)$ reduces to a product of a Hermite polynomial
and a Gaussian:
\begin{equation}
  D_n(z) = 2^{-n/2}\,e^{-z^2/4}\,
            H_n\!\left(\frac{z}{\sqrt{2}}\right) ,
  \qquad n = 0,1,2,\ldots
  \label{eq:PCF_Hermite_app}
\end{equation}
The first few explicit forms are
\begin{align}
  D_0(z) &= e^{-z^2/4} \,,
  \label{eq:D0_app}\\
  D_1(z) &= z\,e^{-z^2/4} \,,
  \label{eq:D1_app}\\
  D_2(z) &= (z^2-1)\,e^{-z^2/4} \,,
  \label{eq:D2_app}\\
  D_3(z) &= (z^3-3z)\,e^{-z^2/4} \,.
  \label{eq:D3_app}
\end{align}
 
\subsection{Recurrence relations}
\label{app:recurrence}
 
The parabolic cylinder functions satisfy the
recurrence relations~\cite{Abramowitz}
\begin{align}
  D_{\nu+1}(z) - z\,D_\nu(z) + \nu\,D_{\nu-1}(z)
  &= 0 \,,
  \label{eq:rec1_app}\\
  D_\nu'(z) + \tfrac{z}{2}\,D_\nu(z)
  - \nu\,D_{\nu-1}(z) &= 0 \,,
  \label{eq:rec2_app}\\
  D_\nu'(z) - \tfrac{z}{2}\,D_\nu(z)
  + D_{\nu+1}(z) &= 0 \,,
  \label{eq:rec3_app}
\end{align}
where primes denote differentiation with respect to $z$.
Relations~\eqref{eq:rec2_app} and~\eqref{eq:rec3_app}
are used in Section~\ref{sec:solution} to derive the
lower spinor component $\psi_2$ from $\psi_1$.
 
\subsection{Wronskian}
\label{app:wronskian}
 
The two linearly independent solutions of~\eqref{eq:Weber_app}
are $D_\nu(z)$ and $D_\nu(-z)$.
Their Wronskian is
\begin{equation}
  W[D_\nu(z),\,D_\nu(-z)]
  \equiv D_\nu(z)\,D_\nu'(-z) - D_\nu'(z)\,D_\nu(-z)
  = -\frac{\sqrt{2\pi}}{\Gamma(-\nu)} \,,
  \label{eq:Wronskian_app}
\end{equation}
which is non-zero for non-integer $\nu$, confirming
linear independence.
This Wronskian enters the delta-function
normalization~\eqref{eq:NE_delta} via the asymptotic
completeness relation for the continuous spectrum.
 
\subsection{Asymptotic behaviour}
\label{app:asymptotics}
 
For large $|z|$ with $|\arg z| < 3\pi/4$~\cite{DLMF}:
\begin{equation}
  D_\nu(z)
  \;\underset{|z|\to\infty}{\sim}\;
  e^{-z^2/4}\,z^\nu
  \left[1 - \frac{\nu(\nu-1)}{2z^2}
          + \mathcal{O}(z^{-4})\right] .
  \label{eq:PCF_asymp_app}
\end{equation}
For the IDMO, the argument is
$z = \xi = \sqrt{2m\omega}\,x$ with complex order
$\nu$ given by~\eqref{eq:nu}.
The exponential factor behaves as
\begin{equation}
  e^{-\xi^2/4}
  = e^{-m\omega x^2/2}
  \quad \text{(real, Gaussian decaying for real } x\text{)},
  \label{eq:PCF_gaussian_app}
\end{equation}
while the prefactor $\xi^\nu$ with
$\mathrm{Im}(\nu) = 1/2$ oscillates, rendering
$\psi_1 \notin L^2(\mathbb{R})$.
On the complex contour $x = re^{i\pi/4}$
(Section~\ref{subsec:contour}), the exponential
becomes $e^{-im\omega r^2/2} \cdot e^{-m\omega r^2/2}$,
which is Gaussian-decaying, restoring normalisability.
 
\subsection{Orthogonality and normalization integral}
\label{app:ortho}
 
For integer orders, the standard orthogonality
on the real line is~\cite{Abramowitz}
\begin{equation}
  \int_{-\infty}^{\infty}
  D_m(z)\,D_n(z)\,dz
  = \sqrt{2\pi}\,n!\,\delta_{mn} \,.
  \label{eq:PCF_ortho_integer_app}
\end{equation}
For the contour normalization of
Section~\ref{subsec:contour}, this gives
\begin{equation}
  \int_{\mathcal{C}} dx\,
  \psi_1^{(m)*}(x)\,\psi_1^{(n)}(x)
  = \mathcal{C}_m^*\,\mathcal{C}_n\,
    \frac{1}{\sqrt{2m\omega}}
    \int_{-\infty}^\infty d\xi\,
    D_m(\xi)\,D_n(\xi)
  = \delta_{mn} \,,
  \label{eq:contour_ortho_app}
\end{equation}
consistent with the normalization
constant~\eqref{eq:psi1_contour_norm}.
 
For general (non-integer, complex) order $\nu$,
the relevant normalization integral for the
delta-function scheme is obtained from the
Wronskian~\eqref{eq:Wronskian_app} and the
asymptotic completeness relation:
\begin{equation}
  \int_{-\infty}^{\infty} dx\,
  \psi_1^*(x;E)\,\psi_1(x;E')
  = \mathcal{N}_E^2\,
    \frac{\sqrt{2\pi}}
         {|\Gamma(-\nu)|}\,
    \delta(E-E') \,,
  \label{eq:delta_norm_app}
\end{equation}
from which the normalization
constant~\eqref{eq:NE_delta} follows upon
using the reflection formula
$|\Gamma(-\nu)|^{-2} = |\Gamma(1+\nu)|^2/\pi^2
\cdot \sin^2(\pi\nu)$ evaluated at
$\nu = (E^2-m^2)/(2m\omega) - 1/2 + i/2$.
 
\subsection{Special value at the origin}
\label{app:origin}
 
The value and derivative of $D_\nu$ at $z = 0$ are
\begin{align}
  D_\nu(0)
  &= \frac{\sqrt{\pi}}{2^{-\nu/2}\,
     \Gamma\!\left(\frac{1-\nu}{2}\right)} \,,
  \label{eq:D_at_0_app}\\
  D_\nu'(0)
  &= -\frac{\sqrt{\pi}}{2^{(-\nu-1)/2}\,
     \Gamma\!\left(-\frac{\nu}{2}\right)} \,.
  \label{eq:Dp_at_0_app}
\end{align}
These are used to verify boundary conditions
and to compute the probability density at $x=0$
for each resonance state.
\bibliographystyle{unsrt}
\bibliography{bib}
\end{document}